\documentclass[submission,copyright,creativecommons]{eptcs}

\usepackage{underscore}       
\usepackage[pdftex]{graphicx}
\usepackage{caption}
\usepackage{subcaption}
\usepackage{amsfonts,amstext}
\usepackage{amssymb,amsmath}
\usepackage{amsthm}

\usepackage{listings}
\usepackage{xcolor}

\definecolor{delimiterColor}{HTML}{B65E47}
\definecolor{numberColor}{HTML}{FF0000}
\definecolor{commentColor}{HTML}{008000}
\definecolor{keyColor}{HTML}{002BFF}

\lstdefinelanguage{maude}
{
	breaklines=true,
	extendedchars=true,
	tabsize=2,
	columns=fullflexible,
	showtabs=false,
	showstringspaces=false,
	showspaces=false,
	showstringspaces=false,
	identifierstyle={\ttfamily},
	keywordstyle={\color{keyColor}},
	ndkeywordstyle={\color{keyColor}},
	stringstyle={\color{delimiterColor}},
	commentstyle={\color{commentColor}},
	ndkeywords={},
	keywords={load, pr, protecting, sort, sorts, op, ops, var, vars,eq, cq, ceq, crl, rl, mb, cmb, endfm, fmod, is, mod, endm, load, =, ==, =/=, ctor, ditto, Object, owise, Oid, prec, assoc, id, if, class, homod, endhom, eof, var, vars, eq, op, ops, pr, inc, protecting, including, ceq, is, tomod, endtom, sort, subsort, subsorts, to, endom, fmod, endfm, mod, endm, endtm, comm, gather, fth, endfth, format, metadata},
	morecomment={[l]{***}},
	morecomment={[l]{---}},
}

\newtheorem{definition}{Definition}

\newtheorem{property}{Property}

\newcommand{\Nat}{\ensuremath{\mathbb{N}}}
\newcommand{\Rat}{\ensuremath{\mathbb{Z}}}

\newcommand{\p}[1]{\ensuremath{\mathtt{p}_{#1}}} 
\newcommand{\tr}[1]{\ensuremath{\mathtt{t}_{#1}}} 

\newcommand{\bag}[1]{\ensuremath{Bag[#1]}}

\title{A Maude Implementation of Rewritable Petri Nets:\\a Feasible Model for Dynamically Reconfigurable Systems}
\author{Lorenzo Capra
\institute{Dipartimento di Informatica}
\institute{Universit\`a degli Studi di Milano, Milan, Italy}
\email{capra@di.unimi.it}
}

\providecommand{\event}{AppFM 2021}
\def\titlerunning{Maude implementation of Rewritable PN}
\def\authorrunning{L. Capra}
\begin{document}
\maketitle

\begin{abstract}
Petri Nets (PN) are a central, theoretically sound model for concurrent or distributed systems but, at least in their classical definition, not expressive enough to represent dynamic reconfiguration capabilities.
On the other side, Rewriting Logic has proved to be a natural semantic framework for several formal models of concurrent/distributed systems. We propose a compact, efficient Maude formalization of dynamically reconfigurable PT nets (with inhibitor arcs), using as a running example the specification of a simple, fault-tolerant manufacturing system. We discuss the advantages of such a combined approach, as well as some concerns that it raises.
\end{abstract}

\section{Introduction}
\label{sec:intro}
Modern distributed systems operate in highly dynamic environments and have to frequently face changing operational conditions. Some system components may become temporarily or permanently unavailable, may appear/disappear, e.g., due to failures and/or dynamic balance policies. Self-adaptation is an effective approach to deal with the increasing complexity and dynamism of such systems. Reconfiguration capabilities are highly required by self-adaptive distributed systems as well as by other types of system, e.g., automated systems, which function without manual intervention, and whose operations are a part of the system control whose processes are completely automated with the help of control loops and special logic.

Formal methods providing the ability to reason about such complex systems are highly demanded. In particular, formal models representing both the system's base structure/functionality and the (usually complex) reconfiguration strategies are extremely important (e.g., to validate early design choices or to verify the system behavior at run-time).

Petri nets (PNs) are a sound, central model of concurrent or distributed systems, but not expressive enough to specify, in a natural way, the ability of some systems to modify their behavior/structure at run-time. Several dynamic PN extensions have been proposed in which the enhanced expressivity is not always supported by adequate analysis techniques. A representative of this category is the “nets within nets” paradigm, introduced by Valk \cite{Valk04}, that gave rise to special High-Level PNs such as \cite{Hornets}. As for PNs with indistinguishable tokens,
we have to mention Reconfigurable PN, a family of PN-based formalisms consisting of a marked net and a separated set of net-transformation rules specified according to the classical, algebraic style of Graph Transformations Systems (as a double pushout) \cite{Ehrig03,Llorens04,reconfnet07,flexnets,Prange2008}. Most research has focused on trying to formulate these models as $\mathcal{M}$-adhesive categories. See \cite{Padberg:2018} for a survey of dynamic PN extensions. 

In this paper, we provide a formalization of ``rewritable'' PT nets with inhibitor arcs (a Turing-complete PN class) in terms of rewriting-logic \cite{rewlog92,rewlog03}, using Maude as a specification language \cite{maude07}. We focus on operational, modelling aspects, by proposing an expressive and efficient framework for the specification/analysis of (automated) distributed systems with reconfiguration capability.
With respect to similar approaches \cite{Barbosa11,RPN-Maude2016},
in which (reconfigurable) PN classes are converted in Maude modules to exploit the model-checking tools of Maude, our encoding provides more data abstraction to ease the modeller task, is more compact and efficient, and promotes the definition of rewrite rules with a higher level of flexibility. Throughout the paper, we use as a benchmark a simple, yet tricky, model of fault-tolerant Manufacturing System. We conduct some experiments of formal verification of properties and briefly discuss of the advantages of using a somehow hybrid modelling approach like that we propose. 

The presented work should be deemed as a preliminary, however, encouraging, step towards a Maude-based tool-set for the specification and the analysis of dynamically reconfigurable PT systems.

\section{PT nets}
\label{sec:backgr}
This section collects a few definitions used in the paper.
We refer to~\cite{ReisigPN}
for a complete description.

\subsection{Multisets}
\label{sec:bag}
A \textit{multiset} (or \textit{bag}) on a domain $D$ is a map $b: D \rightarrow \Nat$, where $b(d)$ is the \emph{multiplicity} of $d$ in $b$. We say 
$d \in b$ if and only if $b(d) > 0$.
We represent a multiset as a weighted ``sum'' of its elements.
$nil_D$ ($nil$, when $D$ is implicit)
denotes the \textit{empty} multiset.
Let $Bag[D]$ denote the set of bags over $D$, and $b_1,b_2 \in Bag[D]$. The \textit{sum} $b_1+b_2$ and the \textit{difference} $b_1-b_2$ are $Bag[D]$ elements such that
$\forall d \in D$:
$b_1+b_2 (d) = b_1(d) + b_2(d)$;  $b_1-b_2 (d) = b_1(d)-b_2(d)$ if $b_1(d) \geq b_2(d)$, $0$ otherwise.
 Note that 
 $-$ is not associative. Similarly, relational bag-operators are defined component-wise:
 e.g., $b_1 < b_2$ if and only if $\forall d \in D:  b_1(d) < b_2(d)$.
With $b_1 <' b_2$ we mean the \emph{restriction} of $<$ to $\{d | d \in b_1 \}$.

\subsection{Place/Transition (PT) Nets with Inhibitor Arcs}
\label{sec:ptnets}
A PT \emph{net} is a 5-tuple $(P,T,I,O,H)$, where:
\begin{description}
\item $P$, $T$ are non-empty, finite sets such that $P \cap T = \emptyset$
\item $I,O,H$ are maps $T \rightarrow \bag{P}$ such that $\forall t \in T: I(t) \neq O(t)$, 
\end{description}

$P$ and $T$ hold the \emph{places} and the \emph{transitions}, respectively. The former, drawn as circles, represent system state-variables, whereas the latter, drawn as bars, represent events causing local \emph{state changes}\footnote{The condition set on transitions ensure this.}. A (distributed) state of a PT net, called \emph{marking}, is defined as a bag $m \in Bag[P]$.

A PT net is a kind of directed, bipartite multi-graph whose nodes are $P$ and $T$. Maps $I$, $O$, $H$ describe the \emph{input}, \emph{output}, and \emph{inhibitor} edges, respectively. Graphically: \includegraphics[height=1.1em]{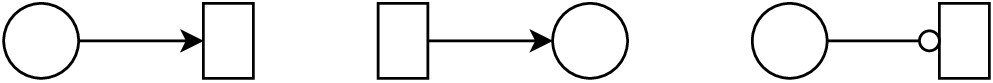}

Let $f \in \{I, O, H\}$: if $k = f(t)(p) > 0$, then a weight-$k$ edge of corresponding type links $p$ to $t$.
The behavior of a PT net is specified by the \emph{firing rule}.
We say that $t\in T$ is \emph{enabled} in a marking $m$ if and only if:
\begin{displaymath}
\ I(t) \leq m \wedge H(t) >' m
\end{displaymath}
If $t$ is enabled in $m$ it may fire, leading to $m^\prime$ (we denote this $m [ t \rangle m'$), where:
\begin{displaymath}
\ m'= m + O(t) - I(t)
\end{displaymath}

We call a pair $(N,m)$, where $N$ is a PT net and $m$ is a marking of $N$, a PT-system.
The interleaving semantics of $(N,m_0)$, where $m_0$ denotes the PT system's initial state, is specified by the \emph{reachability graph} (RG), an edge-labelled, directed graph $(V, E)$  whose nodes are reachable markings. The RG is defined inductively: $m_0 \in V$; if $m \in V$ and
$m [ t\rangle m'$
then $m' \in V$, $m \xrightarrow{t} m' \in E$.

A desirable property of a system $(N,m_0)$ may be the absence of isolated places, i.e., places which are present in $m_0$ but not in $N$. In a highly dynamic context, however, we do not consider this as a strict requirement.   

\section{Maude and Rewriting Logic}
Maude \cite{maude07} is an expressive, strictly declarative language with a sound semantics  in rewriting logic \cite{rewlog92,rewlog03}.
Maude’s statements are (possibly conditional) \emph{equations} (keyword \texttt{eq}) and \emph{rules} (keyword \texttt{rl}).
Both sides of a rule/equation are terms of a given \emph{kind} and may contain typed variables. Both rules and equations have a simple rewriting semantics in which instances of the lefthand side pattern are replaced by corresponding instances of the righthand side.

A Maude \emph{functional} module (keyword \texttt{fmod}) contains only equations and is a functional program defining one or more operations through equations, used as simplifications.
A Maude \emph{system module} (keyword \texttt{mod}) contains rules and possibly equations. Rules are also computed by rewriting terms from left to right, but represent local transitions in a (concurrent) system. Although declarative in style and with a clear logical semantics, system modules are non-functional.

In Maude, a distributed system's state is usually represented as a kind of associative ``multiset''.
Rules apply \emph{concurrently} to different portions of a system, leading to a new state.
Unlike for equations, there is no assumption on the confluence of rewrites. 
Reactive systems may not even have any final states.

Maude features expressivity, simplicity and performance.
A wide range of systems are naturally expressible, from sequential, deterministic to highly concurrent non-deterministic ones.
Maude may be used as a formal specification language, as a programming language and as a
meta-language in which other formalisms, languages and logics can be naturally expressed.
In this paper, we use Maude to specify adaptable PT systems and  their operational semantics.

Maude's expressivity is achieved through: equational pattern matching modulo operator equational attributes;
user-definable operator syntax/evaluation strategy; sub-typing(sorting) and partiality (kinds); generic types; reflection.

A Maude program is a logical theory, and a Maude computation is a deduction according to the axioms specified in the program. Under certain executability conditions of modules, the mathematical and the operational semantics coincide. We refer to \cite{membeqlog00} and \cite{sysmodsem01}  for functional and system modules, respectively.

A functional module specifies an \emph{equational theory} in membership equational logic \cite{membeqlog98,membeqlog00}.
Formally, such a theory is a pair $(\Sigma,E \cup A)$, where $\Sigma$ is the signature, that is, the specification of all the sort, subsort, kind\footnote{A \emph{kind} is an implicit equivalence class gathering all sorts connected by the subsort partial-order relation; terms having a kind but not a sort may be considered as \emph{undefined} or \emph{errors}.}, and (overloaded) operator declarations in a functional module (plus the imported  functional modules if any);
$E$ is the collection of (conditional) equations and memberships declared in the module(s), and $A$ is the collection of equational attributes (\texttt{assoc}, \texttt{comm}, and so on) of operators (treated as predefined equations).

The models underlying functional modules are algebras, i.e., sets of data with related operations. For example, the model for (instantiated generic) module \texttt{BAG[A]} is the multisets over \texttt{A} with their standard operations. 
The family of $\Sigma$-ground terms $T_{\Sigma}$ defines a model called $\Sigma$-algebra ($T_{\Sigma(X)}$ denotes the whole set of terms).
According to Goguen and Burstall \cite{Burstall1982}, the best model of $(\Sigma,E \cup A)$ is one that satisfies $E \cup A$ and is both junk-free (all elements can be denoted by $T_{\Sigma}$ terms) and confusion-free (only elements that are forced to be equal by $E \cup A$ are identified). This model, called the \emph{initial algebra} of $E \cup A$ and denoted $T_{\Sigma /E \cup A}$, does exist \cite{membeqlog00} and provides the denotational semantics of the Maude functional module specifying $(\Sigma,E \cup A)$.
Formally, $T_{\Sigma /E \cup A}$ is the quotient of $T_{\Sigma}$ in which the equivalence classes hold terms that prove equal using $E \cup A$.

If the axioms $E$ are Church-Rosser and terminating modulo $A$ (each ground term is thus simplified in a unique way regardless of the order in which equations apply)
there is an intuitive, equivalent description for $T_{\Sigma /E \cup A}$.
The final values (\emph{canonical} forms) of all ground terms form an algebra called the \emph{canonical term algebra}, denoted $CAN_{\Sigma /E \cup A}$.
By definition, the \texttt{reduce} command of Maude interpreter reduces operators to their values in this algebra.
The coincidence of the denotational and operational semantics is expressed by $T_{\Sigma /E \cup A} \cong CAN_{\Sigma /E \cup A}$.

A Maude's system module, including the imported submodules,
specifies a generalized \emph{rewrite theory} \cite{rewlog92,rewlog03},
that is, a four-tuple $\mathcal{R}= (\Sigma,E \cup A,\phi,R)$
where $(\Sigma,E \cup A)$ is the membership equational theory specified by the signature, equational attributes, and equation statements in the module; $\phi$ is a map specifying, for each operator in $\Sigma$, its frozen arguments; and $R$ is a set of rewrite rules
\footnote{$R$ rules don't apply to frozen arguments; in the paper we do not use frozen arguments.}.

Intuitively, a rewrite theory specifies a concurrent system. The equational part $(\Sigma,E \cup A)$ specifies
the algebraic structure of the states, formalized by the initial algebra $T_{\Sigma /E \cup A}$. The rules $R$ (and $\phi$) specify the system’s dynamics, that is, the possible concurrent transitions of the system.
In our context, they represent changes to the state/structure of a PT system.
In rewriting logic, concurrent transitions become rewrite proofs; since several proofs may correspond to the same computation (because of different, equivalent interleavings), rewriting logic has an equational theory of proof equivalence \cite{rewlog03,rewlog92}.

The initial model $\mathcal{T}_{\mathcal{R}}$ of $\mathcal{R}$ associates to each kind $k$ a labeled transition system (a category) whose states are $T_{\Sigma /E \cup A,k}$, and whose
transitions take the form: $ [t] \overset{[\alpha] }{\rightarrow} [t']$, with $[t],[t'] \in T_{\Sigma /E \cup A,k}$, and $ [\alpha]$ an equivalence class of rewrites modulo the equational theory of proof-equivalence. Different $ [\alpha]$ represent different “truly concurrent” computations of the system specified by $\mathcal{R}$.

The executability conditions for system modules match the notion of ground \emph{coherence} between rules and equations \cite{sysmodsem01}. Assuming that $E \cup A$ is Church-Rosser and terminating, an efficient strategy (the one adopted by the Maude \texttt{rewrite} command) is to first apply the equations to get a canonical form, then apply a rewrite rule in $R$. Coherence ensures that this strategy is complete, i.e., any rewrite of $t \in T_{\Sigma}$ with $R$ is also possible with $t$’s canonical form.
Coherence is crucial because rewriting modulo an equational theory is in general undecidable. It reduces rewriting with $R$ modulo $E \cup A$ to rewriting with $E$ and $R$ modulo $A$, which is decidable given an $A$-matching algorithm.

Checking the confluence/termination and coherence properties of Maude modules is under the user’s responsibility. In many cases, it is possible to prove these properties using the tools of Maude Formal Environment (MFE), available at \url{https://github.com/maude-team/MFE}, most of which are currently integrated with the Maude's interpreter. The Maude modules listed in this paper have been proven to be executable. Their functional part also meets two other desirable properties: i) each (equivalence class of a) ground term has a least sort (kind); ii) the canonical form of a well-defined ground term is only built of constructors (characterized by the \verb|ctor| operator attribute). A syntactical condition (called term pre-regularity) and Maude’s Sufficient Completeness Checker (SCC) have been used to verify properties i) and ii), respectively.

\section{Running example:  a reconfigurable, fault-tolerant MS}
\label{sec:exe}
As an example of ``rewritable'' PT system, we use the model of a manufacturing system (MS) equipped with dynamic reconfiguration capabilities to face failures affecting the production lines. Despite its simplicity, this is an interesting benchmark for any formalism intended to specify highly dynamic systems. We use a small variant (as for the reconfiguration steps) of the model introduced in~\cite{CamilliECSA2018}.

\begin{figure}[ht]
    \centering
    \includegraphics[width=\textwidth]{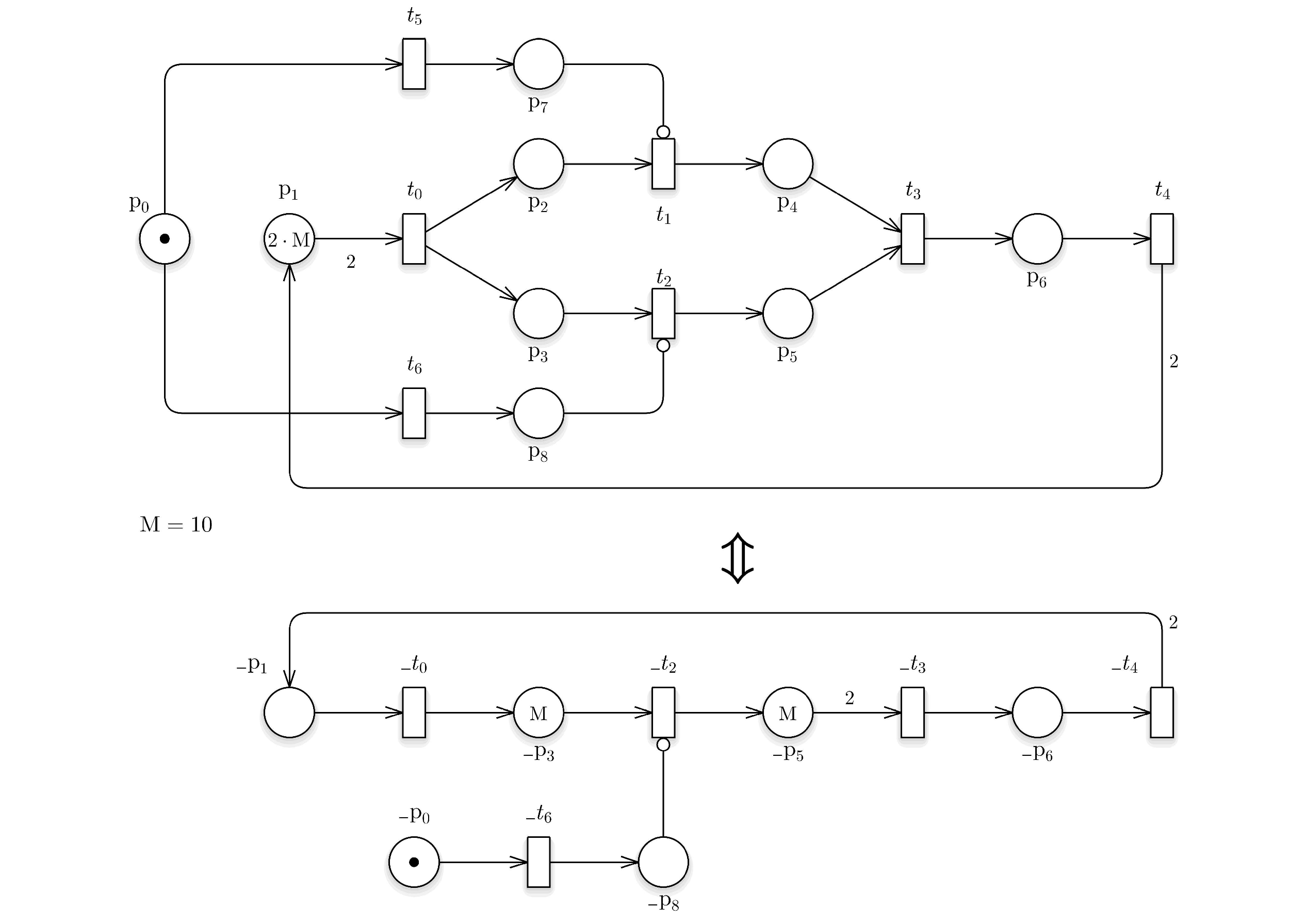}
    \caption{The MS and its adaptation upon a fault on line 1 (a fault on line 2 is managed analogously).}
    \label{fig:FMS}
\end{figure}

The MS (Fig.~\ref{fig:FMS}, top) is composed of two symmetric production lines that (globally) work an even number of raw pieces. Pairs of worked pieces are assembled to get the final artefacts.
Either line may get broken occasionally,
in which case the system reconfigures itself so that it can continue working using the line left. The reconfiguration involves
some changes to the system's topology and the transfer of residual raw pieces staying on the faulty component to the available one (Fig.~\ref{fig:FMS}, bottom).
If another fault affects the reconfigured system then (after a hypothetical repair phase, not represented) the MS goes back to its nominal configuration.
The model's parameter $M \in \Nat^+$ defines the number of raw pieces ($2 \cdot M$) worked during an entire production cycle. Each final artefact is immediately replaced by a pair of pieces to be worked, what makes the system's nominal behaviour cyclic.

Tokens flowing through the PT places represent either raw or worked pieces.
The two production lines are modelled by the symmetrical subnets $\{$ $\mathtt{p_2}$, $\mathtt{t_1}$  $\mathtt{p_4}$ $\}$ (line 1) and $\{$ $\mathtt{p_3}$, $\mathtt{t_2}$, $\mathtt{p_5}$ $\}$ (line 2).
The assembly of raw pieces is represented by transition $\mathtt{t_3}$.
Transition $\mathtt{t_0}$ models the loader component, that (initially) picks up two raw pieces at a time from a storage represented by place $\mathtt{p_1}$
putting them onto the lines.
Transition $\mathtt{t_4}$ models the immediate reload of the MS.
The model includes also a very simple specification of fault occurrence,
represented by transitions $\mathtt{t_5}$ (for line 1) and $\mathtt{t_6}$ (line 2).
A fault on a line causes its immediate block, which is modelled, for example, by the inhibitor arc linking $\mathtt{p_7}$ to $\mathtt{t_1}$.
We assume that a simultaneous failure of both lines is not possible.
For this reason, the transitions $\mathtt{t_5}$ and $\mathtt{t_6}$ are in symmetric structural conflict.

The PT net at the bottom of Fig.~\ref{fig:FMS} clearly illustrates the changes to the MS layout that are required so that it can work using the only available line upon a fault, preserving the expected system behaviour. The marking of the PT system at the top represents the initial state, whereas that of the PT system at the bottom reflects the situation just after a fault occurrence: we may imagine (as explained later on) that $M$ residual raw pieces on line 1 (place $\mathtt{p_2})$ have been moved to line 2 (place $\mathtt{p_3})$.
Two major concerns that will be addressed in the next section are how to formally specify the MS reconfiguration(s) described in Fig.~\ref{fig:FMS} and the condition(s) under which it occurs.  

Although PT nets with inhibitor arcs are a Turing-powerful formalism, their limited expressivity doesn't help in representing scenarios like that described (unless you enormously complicate the model). In the rest of the paper, we show how Maude can be used as a formal framework (in combination with PT) to overcome these drawbacks.

\section{Encoding Rewritable PT systems in Maude}
In this section, we review the Maude formalization of PT systems with dynamic reconfiguration capabilities. The full list of Maude source files is available at \url{https://github.com/lgcapra/rewpt}. Only one (system) module refers to a specific model, all the others should be seen as reusable components. The reusable part is built of a set of functional modules specifying the PT system signature and some structural equivalences (the signature's main module is \verb|PT-SYS|) and a system module (\verb|PT-EMU|) which describes the system dynamics (firing rule). The main Maude modules are listed in the Appendix A. 

The Maude formalization relies on two generic functional modules, \verb|BAG{X}| and \verb|MAP+{X,Y}|, listed in the Appendix B.
The parameters of these generic types (all satisfying the built-in Maude trivial theory \verb|TRIV|\footnote{Maude uses views to instantiate the formal type-parameters (Maude theories) of a generic module to concrete types (modules). We only report an example of theory and view, because they are very intuitive.}) correspond to the bag's support and to the map's domain and codomain.

Differently from \cite{RWLPN2001,RPN-Maude2016}, bags on a set are not merely represented as the free commutative monoid on that set. Despite its elegance and simplicity, such a formalization is unsuitable for bags with many repetitions. 
Our formalization is more convenient and practical, providing the $\{$\verb|_._|, \verb|_+_|, \verb|_[_]| \verb|_-_|, \verb|_<=_|, \verb|_>'_|, \verb|set|$\}$ operators (the first two \emph{constructors}). In particular, the overloaded, commutative/associative sum makes it possible to intuitively represent a bag in a compact way, e.g., \verb|3 . a  + 10 . b|. Note the use of Maude advanced features, such as operator equational attributes, precedence and gathering (a term like \verb|3 . a  - 1 . a + 1 . a| is parsed as: \verb|3 . a  - (1 . a + 1 . a)| $\equiv$ \verb|1 . a|), and the \verb|otherwise| equation attribute, which permits a straightforward, efficient definition of difference, relational and lookup (\verb|_[_]|) operators. The \verb|set| operator sets up a multiplicity  for an element overwriting the older one. Setting a zero-multiplicity corresponds to removing an element.

The module \texttt{MAP+} defines a map as a “set”  of entries, built with the associative/commutative operator \verb|_,_|. Sort \verb|Entry|, whose only constructor is the operator \verb"_|->_", is a subsort of \texttt{Map}.
The constant \verb|undefined| of the kind \verb"[Y$Elt]" (corresponding to the sort \verb"Y$Elt") represents an undefined association.
The overloaded operator \verb|insert| adds a new key-value pair to a map, consistently updating the map in the event the key is already present. Notice the use of matching and of the \verb|otherwise| attribute to distinguish these two cases easily. The auxiliary operation \verb|hasMapping| checks whether a key actually has an associated entry in a map. The lookup operator is represented by \verb|_[_]|.
With respect to the built-in \texttt{MAP} module of Maude, \texttt{MAP+} has two significant extensions: as in the built-in module, the constructor \verb|_,_| doesn't guarantee (for the sake of efficiency) the uniqueness of keys in a map; a conditional membership equation (using an auxiliary predicate), however, assigns a term of kind \verb|[Map]| the sort \verb|Map| only if it doesn't contain duplicated keys. The overloaded operator \verb|remove|, instead, withdraws a given key from the map.
We have defined bags apart, rather than as particular maps, for efficiency and readability.

\paragraph{PT system signature.} The Maude signature of PT systems strictly follows the definition given in Sect.~\ref{sec:backgr}, so that passing from a formalization to the other is straightforward.
The nodes of a PT net are trivially defined (see module \texttt{TRAN}) as terms with a subscript, e.g., \verb|t(1)|. The \texttt{PLACE} module is analogously defined. A richer definition is possible, as discussed later. The module \texttt{IMATRIX} provides a compact representation for the incidence matrix of a PT net transition, syntactically represented as a triplet (according to the three types of edges) of bags on the set of places through the \verb|[_,_,_]| constructor. \texttt{IMATRIX} imports \texttt{BAG} in a protected way (i.e., without introducing junks/confusion) by renaming some sorts/operators for readability. Two auxiliary operators are defined, to \verb!remove! a given place from the transition incidence matrix and to check the adjacency of the transition to a place (the predicate \verb!in!).

The central module \texttt{PT-SYS} defines the PT system signature. It builds on \texttt{MAP+} and \texttt{IMATRIX}, both imported in a protected way, with some sort/operator renaming. In particular the parameterized sort \verb|Map{Tran,Imatrix}| is renamed \verb|Net|. That is, a PT net is syntactically expressed as a set of entries built of transitions and (local) incidence matrices. This formalization is very intuitive and allows structural changes to a system to be efficiently and consistently specified. A PT system term is a simple juxtaposition (empty constructor \verb|__|) of a \verb|Net| term and a \verb|BagP| term (the latter representing a PT net marking).
For convenience, the module provides three operators on the sort \verb|Net| that correspond to the maps $I, O, H$ of the PT net formal definition, and three predicates matching well-known conditions: \verb|enabled| checks for the enabling condition of a PT net transition in a certain marking; \verb|dead| (which builds on \verb|enabled|) checks that a PT system is dead, i.e., no transition is enabled in the associated marking; \verb|in| checks that a place belongs to a PT system (to the net and/or the associated marking). Note that the co-domains of \verb|__|, \verb|enabled|, and \verb|In|, \verb|Out|, \verb|Inh| are \emph{kinds} (\verb|[System]|, \verb|[BagP]|, and \verb|[Bool]|, respectively), i.e., these are somehow \emph{partial} operations. As for \verb|__|, the reason is that the net of a PT system must not be empty (constant \verb|emptyN|): a membership axiom specifies those terms having sort \verb|System|. As for the others, the subtle reason is that they are defined only for \verb|Tran| operands belonging to the specified PT net (system).
Thus, \verb|enabled(net m0,t(1))| results in \verb|false| if
\verb|t(1)| belongs to \verb|net| and is not enabled in \verb|m0|.

The module \texttt{PT-SYS} includes two equations specifying structural equivalences between PT nets. One eliminates transitions having a null effect, according to the definition in Sect.~\ref{sec:backgr}. The other eliminates transitions structurally dead, because linked to a place by an input edge and by an inhibitor edge of non lesser weight. The commented equation would remove isolated places from a PT system (as we said, this is not mandatory).

\paragraph{PT system dynamics.} The generic (system) module \texttt{PT-EMU} specifies the PT system operational semantics. The module's parameter must satisfy the Maude theory \texttt{PTSYSTH}, that requires two zero-arity operators describing a marking and a net. A conditional rewrite rule, syntactically much like the definition in Sect.~\ref{sec:backgr}, specifies the PT system firing rule. Notice the use (in the rule's condition) of the conjunction associative connective \verb|/\| and of a matching equation (\verb|t := t’|) which makes the rule compact and efficient to apply. The free variables \verb|I|,  \verb|O|, \verb|H|, \verb|N'| (two of which used in the rule's right-hand side) are instantiated by matching the left-hand side of the matching equation against the canonical ground term bound to variable \verb|N| (that occurs on the rule's left-hand side).

\paragraph{Rewritable PT systems.} The model-dependent part consists of a single Maude (system) module importing \texttt{PT-SYS} and satisfying the Maude theory \texttt{PTSYSTH}, which is (in part) mechanically derived from a PT system $(N,m_0)$. The module \texttt{RWPT-FMS}, e.g., comes from the PT system in Fig.~\ref{fig:FMS}-top, describing the FMS nominal behavior. Two equations assign terms \verb|net|, \verb|m0| the expressions encoding the PT net $N$ and the initial marking $m_0$, respectively. These expressions are obtained by arbitrarily defining two bijections $\phi_P : P \rightarrow \{0,\ldots,|P|-1\}$, $\phi_T : T \rightarrow \{0,\ldots,|T|-1\}$ (in our example, implicitly defined by subscripts of PT nodes). For the rest, the encoding of $N$ (which consists of directly encoding the functions $I, O, H$ into a \verb|net| term, i.e., a set of entries \verb|T:Tran -> Q:Imatrix| ) and $m_0$ (as a \verb|BagP| term) turns out to be straightforward (we skip the details). 

Note the use of zero-arity, sort \verb|Nat| operator \verb|M| in \texttt{RWPT-FMS}, that allows the module to be syntactically parametric in the initial marking of place $p_1$.

The remaining part of the module contains a (possibly empty) set $R$ of rewrite rules specifying structural/state changes that may occur during
the evolution of the system initially matched by the value of term \verb|net m0|. With respect to similar approaches \cite{RPN-Maude2016}, where rewrite rules strictly follow, both syntactically and semantically, the paradigm of algebraic GTS (based, e.g., on DPO rules) and require quite complex consistency checks (e.g., to verify the respect of gluing conditions), our Maude-based specification fosters a much more flexible, free-style, though rigorous, formalization of (rule-based) system transformations. 

We can classify rewrite rules according to the kind of $T_{\Sigma(X),k}$ terms in rule left- and right-hand sides: \verb|Place|, \verb|Tran|, \verb|BagP|, \verb|Imatrix|, \verb|Net|, \verb|System|. Except for rules of \verb|System| type, all the other rules are local, that is, they act on one or more portions of a term describing a PT system. This is perfectly coherent with the operational semantics, intrinsically distributed, of rewriting logic, ensuring a lot of flexibility, but may have undesirable effects. A rule of \verb|BagP| type, e.g., may touch both the marking and some local incidence matrix of a \verb|System|.

The system module \texttt{RWPT-FMS} specifies two \verb|System|-level (conditional) rewrite rules that model the system transformations
described in Fig.~\ref{fig:FMS}. Both rule \verb|r1| (nominal\verb|=>|faulty) and
rule \verb|r2| (faulty\verb|=>|nominal) are \emph{symmetric}, i.e., refer to any of the two lines. In other words, each rule folds two ordinary rules. In rule \verb|r1|, this is achieved by the matching of rule's left-hand pattern(s), which contains some variables (by convention, denoted by capital letters): for example, variable \verb|P3| is instantiated to either ground value \verb|p(2)| or \verb|p(3)|, depending on whether a fault occurred on line 1 or line 2 (the opposite, as for variable \verb|P2|). In rule \verb|r2|,
the two matching equations in the rule's condition make the free variables \verb|P2|, \verb|P3| be instantiated one to \verb|p(2)| and the other to \verb|p(3)|, depending on which line is broken. The condition of rule \verb|r1| ensures that the transformation nominal\verb|=>|faulty may take place when there are no residual ``worked'' pieces on the faulty line awaiting the assembly. Residual raw pieces on that line are then contextually moved to the working line (the marking of place bound to \verb|P3| is cleared and that of place bound to \verb|P2| increased correspondingly).
The last clause of rule \verb|r2|'s condition makes the back-transformation faulty\verb|=>|nominal
may take place when the system (which operates in a degraded way) enters a deadlock, upon another fault breaking the only line left: this happens when in the (broken) line there is at most one piece ``worked'' and all the other pieces still raw (the total of pieces is $2 \cdot M$). In order for the system to safely restart (after a global repair), $M$ raw pieces are moved from that line 
to the one newly (re-)introduced ($M$ tokens are withdrawn from the place bound to \verb|P2| and added to the place bound to \verb|P3|).

\paragraph{A library of base transformations.} In the running example, we use two monolithic, non-trivial rewrite rules, each implementing a transformation of the system as a whole. Beside being complex, such an approach may be impractical when dealing with realistic adaptive systems, where changes are gradual, local and follow a strategy (procedure). Using our formal framework, you can define rewrite rules of varying granularity and locality, with both simplicity and great flexibility. To further ease the modeling task, however, we provide a minimal (extensible) set of base net-transformations specified by simple operators. The (functional) module \texttt{PT-RWLIB} collects two main operators (others are implicitly provided by modules \verb|BAG|, \verb|MAP+|, \verb|IMATRIX|). \verb|setw| allows one to update the weight of any edge in a PT net. For convenience, the edge type is specified as fourth argument of the operator: three constants of sort \verb|Atype| are available for that. New edges/nodes may be added to a net using \verb|setw|. Edges may be also removed with \verb|setw|, using a zero-weight. This may indirectly cause the erasure of nodes that removed edges are incident to. The operator \verb|remove| withdraws a place from a net. The homonym operator of \verb|MAP+| may be used to remove a transition from a net. Using \verb|BAG| operators (e.g., \verb|set|) we may freely modify the marking of a PT system and (local) incidence matrices, add/remove places, and so on. Module \texttt{PT-RWLIB} defines also an operator \verb|w| to easily introspect the edge weight.
This operator implements a partial function, as some of the operators of module \texttt{PT-SYS}. 

For example, assume that the MS switches from one of the configurations shown in Fig.~\ref{fig:FMS} to the other as a result of a sequence of small, local changes to the system's structure/state, carried out upon a fault occurrence. In this scenario,
a thing to eventually do is to stop loading raw pieces into the MS. This is implemented by the (commented) rewrite rule \verb|r3| in \texttt{RWPT-FMS} (requiring module \texttt{PT-RWLIB}). Using the operator \verb|setw|, a new empty input place (\verb|p(9)|) is linked to transition \verb|t(0)| (the loader) to temporarily disable it.
The rule's condition ensures that place \verb|p(0)| is empty (i.e., a fault has occurred) and \verb|p(9)| is not yet in the system. This rewrite step may be used in both the transformations described in Fig.~\ref{fig:FMS}.

\subsection{Base notions and properties of rewritable PT systems}
\label{sec:proprty}
In this section, we provide an essential theoretical basis for the Maude formalization of rewritable PT systems through a few intuitive notions and properties. We refer to the \emph{canonical} form of ground terms,
that does exist and, if a term is well-defined (i.e., has an associated least \emph{sort}), is built of constructors, since Maude modules satisfy the executability conditions (end of Section \ref{sec:backgr}). 

\begin{property}[correspondence between PT systems and well-defined terms]
A PT system $S = (N, m)$ has an associated ground term of sort \verb|System|, vice-versa, a ground term of sort \verb|System|
represents a PT system (up to isomorphism\footnote{$S$ and $S'$ are isomorphic iff there are a two bijections $\phi_p: P \rightarrow P'$, $\phi_t: T \rightarrow T'$, preserving the edges and the initial markings.}).
\label{prop:PTsys-term}
\end{property}
We have described how to get a \verb|System| term from $S = (N, m)$.
Vice versa, we observe that a canonical term of sort \verb|System| is built (by definition of \texttt{PT-SYS} and \texttt{MAP+}) of \verb|(n m)|, with \verb|n| being a non-empty set of entries \verb|T:Tran -> Q:Imatrix| (without duplicate-keys) with the first two components (in, out) of each local incidence matrix unequal, and $m$ a bag of places. By the way, the property above sets a bijection, denoted $\psi$ (a pair $\psi_P, \psi_T$), between (classes of isomorphic) PT systems and well-defined \verb|System| terms.

Let $r$ be a rewrite rule, $t$, $t'$ two ground terms of kind $k$. The notation $t \overset{r(\sigma)}{\rightarrow} t'$ means that 1) the rule's lefthand side $u \in T_{\Sigma(X),k}$
matches $t$ (i.e., there is a ground substitution $\sigma$ such that $\sigma(u) = t$)\footnote{$\sigma$ nay be empty is $u$ is a ground term; if $r$ is a conditional rule $\sigma$ may involve free variables introduced by matching equations used in the rule's condition.}, 2) $t$ is rewritten to $t'$ using $r$, $\sigma$.

Given a PT system $S = (N, m_0)$, let \texttt{RWPT-}$S$ represent a system module (satisfying \texttt{PTSYSTH}) in which the term \verb|(net m0)| encodes $S$, $R$ be the set of rewrite rules defined in \texttt{RWPT-}$S$, and $S$\verb|-EMU| the module \verb|PT-EMU| whose parameter is instantiated (via an obvious view) to \texttt{RWPT-}$S$.

The interleaving semantics of a rewritable PT system specified by module \texttt{RWPT-}$S$ is expressed naturally by the labelled transition system, denoted $RWLT_S$, which is built from the initial term/state \verb|(net m0)|.

\begin{definition}[State-transition system of \texttt{RWPT-}$S$]
Let $R' = R \cup \{\verb|firing|\}$. $RWLT_S$ is an edge-labelled, directed graph $(V_{RW_S}, E_{RW_S})$ inductively defined:\\
$\verb|(net m0)| \in V_{RW_S}$;
if $s \in V_{RW_S}$ and $s \overset{r(\sigma)}{\rightarrow} s'$ then: $s' \in V_{RW_S}$, $s \overset{r(\sigma)}{\rightarrow} s' \in E_{RW_S}$.
\label{def:lts}
\end{definition}

By default, the Maude interpreter's \texttt{search} command explores the state-space associated with an initial term by executing one-step rewrite rules in a fair, breadth-first way, therefore, coherent with the definition above.

\texttt{RWPT-}$S$ includes the ordinary behaviour of the PT system $S$.
\begin{property}[RG inclusion]
\label{prop:RGinclusion}
$RWLT_S$ contains a sub-graph isomorphic to $RG_S$.
\end{property}
\vspace{4pt}
\noindent It directly follows from the fact that, by definition, for any transition $t$ and for any marking $m$ of $S$: $m [t \rangle m'$ if and only if $\psi_P(m) \overset{\texttt{firing}(\sigma)}{\longrightarrow} \psi_P(m')$, with $\sigma(\verb|T|) = \psi_T(t)$ (\verb|T| is the variable used in rule \verb|firing| of module \verb|PT-EMU|). 

Notice that, in the event of badly defined/used rules, we may reach undefined (error) states, despite the initial system \verb|(net m0)| is well-defined.
 For example, the rule
 
{\small\verb"crl  : (N, T |-> [I, 1 . P, nilP], T' |-> [1 . P, O, nilP]) S  =>"

\verb"(N, T |-> [ I, O, nilP]) S if I[P] = 0 /\ O[P] = 0 /\ S[P] = 0 ."}

\noindent that aggregates two transitions connected by an intermediate empty place not linked to any other transitions, rewrites the  \verb|System| ground term

{\small\verb"(t(1) |-> [1 . p(1),1 . p(2),nilP], t(2) |-> [1 . p(2),1 . p(1),nilP]) nilP"}

\noindent into:
\verb|(emptyN nilP)|, an undefined term of kind \verb|[System]| (due to the equations of functional module \texttt{PT-SYS}).

\begin{definition}[Well-defined \texttt{RWPT-}$S$]
\label{def:welldef-lts}
\texttt{RWPT-}$S$ specifies a well-defined rewritable PT system if and only if all reachable states in $V_{RW_S}$ are terms of sort \verb|System|.
\end{definition}

\paragraph{Rule validation.} 
The module \texttt{RWPT-FMS} is well-defined. In general, however, ensuring the well-definiteness of Maude system modules specifying rewritable systems may not be simple. There are two approaches, shortly discussed in the following, each with different possible implementations.

One consists of defining (structurally) valid rewrite rules, and works also in the event the system state-space is infinite.

\begin{definition}[Valid rewrite rule]
\label{def:validr}
$r \in R$ is valid if and only if, for any ground term $s$ of sort \verb|System|, if $s \overset{r(\sigma)}{\rightarrow} s'$ then $s'$ is of sort \verb|System|.
\end{definition}

\vspace{5pt}
Each Maude rewrite rule \verb|crl [r] : s => s' if cond|, where \verb|cond| is the (possibly empty) rule's condition and \verb|s|, \verb|s'| $\in T_{\Sigma(X),\verb|System|}$, may be rephrased as a valid rule using the built-in \emph{sort predicate}
\begin{center}
\verb|crl [vr]  s => s' if cond /\ s' :: System|     
\end{center}

The weak spot of this elegant and efficient solution is that it may shadow bad design choices. As an alternative, we might define rewrite rules exclusively composed of \emph{safe} net-operators. For example, the operator \verb|setw|, defined in module \texttt{PT-RWLIB}, always results in a term of sort \verb|Net|, possibly \verb|emptyN| (for the sake of flexibility). The operator \verb|setwS|, which builds on  \verb|setw|, also guarantees that the resulting term is a non-empty PT net (note the mixed use of \verb|owise| equation attribute and a matching equation).

\section{Property Verification}
In this section, we briefly address some tools that are available to formally verify the properties of a Maude-based specification of rewritble PT systems, and we present a few experimental data. The goal is not to contend with well-established model-checkers based on PN and/or related formalisms, but to show some potential benefits of the methodology proposed to specify dynamically reconfigurable systems.

A Maude system module (which specifies a rewrite theory) provides an executable formal model of a distributed system. Under appropriate conditions, we can check that this model satisfies some properties, or obtain counterexamples. This kind of model-checking analysis is quite general and builds on the \verb|search| command, which allows one to explore (following a breadth-first strategy) the reachable state-space in different ways. For example, using bounded model-checking, if the system state-space is huge (or even infinite), or model-checking of infinite-state systems through abstractions.
Under finite reachability assumptions, we might more efficiently model check any linear time temporal logic (LTL) property of a system module using the LTL Maude modules. We here focus on a simple, yet helpful, model-checking capability, namely, the model checking of invariants using the \verb|search| command.

One invariant we might like to verify about a rewritable PT system is deadlock freedom. A straightforward way to check this property is to issue the search command below, which searches for \emph{any} final states of our running example starting from the initial PT system configuration.

\verb|Maude> search in FMS-EMU : net m0 =>! X:System .|

\noindent It gives no solution, meaning that the self-adaptive MS is deadlock-free\footnote{Using the MAUDE LTL modules we might even check that the initial state is a home-state, i.e., the system is cyclic.}. Table \ref{tab:perf} reports some data about the performance of this command, as the system size varies. The experiments were carried out on a Intel Core i7-6700 equipped with 32 GB of RAM.

Another interesting search concerns the existence of dead states inside the different configurations (the nominal one and the two symmetric, faulty ones) that the system enters during its evolution. The command to issue is ( \verb|*| means ``in zero or more steps", \verb|dead| is the predicate defined in module \verb|PT-SYS|):
 
\verb|Maude> search in FMS-EMU : net m0 =>* X:System such that dead(X:System) .|

\noindent There are six solutions (for any $M$), two for each configuration of the system. For example, for $M = 50$, we get (for readability, in the following excerpt of the command's output, we use the term \verb|net| to denote the system's nominal configuration instead of the much longer canonical form)

\begin{verbatim}
Solution 1 (state 133541)
states: 133542  rewrites: 16647633 in 65130ms cpu
X:System --> (t(0) |-> [1 . p(1),1 . p(3),nilP], t(2) |-> [1 . p(3),1 . p(5),
1 . p(8)], t(3) |-> [2 . p(5),1 . p(6), nilP], t(4) |-> [1 . p(6),2 . p(1),
nilP], t(6) |-> [1 . p(0),1 . p(8),nilP]) 1 . p(8) + 100 . p(3)

Solution 2 (state 133542)
states: 133543  rewrites: 16647811 in 65140ms cpu 
X:System --> (t(0) |-> [1 . p(1),1 . p(2),nilP], t(1) |-> [1 . p(2),1 . p(4),
1 . p(7)], t(3) |-> [2 . p(4),1 . p(6), nilP], t(4) |-> [1 . p(6),2 . p(1),
nilP], t(5) |-> [1 . p(0),1 . p(7),nilP]) 1 . p(7) + 100 . p(2)

Solution 3 (state 142860)
states: 142861  rewrites: 17841539 in 68140ms cpu 
X:System --> (t(0) |-> [1 . p(1),1 . p(2),nilP], t(1) |-> [1 . p(2),1 . p(4),
1 . p(7)], t(3) |-> [2 . p(4),1 . p(6), nilP], t(4) |-> [1 . p(6),2 . p(1),
nilP], t(5) |-> [1 . p(0),1 . p(7),nilP]) 1 . p(4) + 1 . p(7) + 99 . p(2)

Solution 4 (state 142865)
states: 142866  rewrites: 17842109 in 68140ms cpu
X:System --> (t(0) |-> [1 . p(1),1 . p(3),nilP], t(2) |-> [1 . p(3),1 . p(5),
1 . p(8)], t(3) |-> [2 . p(5),1 . p(6),nilP], t(4) |-> [1 . p(6),2 . p(1),
nilP], t(6) |-> [1 . p(0),1 . p(8),nilP]) 1 . p(5) + 1 . p(8) + 99 . p(3)

Solution 5 (state 1083022)
states: 1083023  rewrites: 143322169 in 781220ms cpu
X:System --> net 1 . p(8) + 50 . p(3) + 50 . p(4)

Solution 6 (state 1084477)
states: 1084478  rewrites: 143524908 in 782130ms cpu
X:System --> net 1 . p(7) + 50 . p(2) + 50 . p(5)
\end{verbatim}

We may suppose a generic form for the six states above in which expressions $M$, $2\cdot M$, $2\cdot M - 1$ should be used in place of values 50, 100, 99, respectively. 

As a last example of use of \texttt{search}, we check whether the system correctly evolves from an inner deadlock: the command below lists the \verb|system| states reachable in one-step from a dead state which refers to the system's nominal configuration after line 2 gets faulty (for $M = 10$). 

\verb|Maude> search net  1 . p(8) + 10 . p(3) + 10 . p(4) =>1 X:System .|
 
\begin{verbatim}
Solution 1 (state 1)
states: 2  rewrites: 23 in 0ms cpu
X:System --> (t(0) |-> [1 . p(1),1 . p(2),nilP], t(1) |-> [1 . p(2),1 . p(4),
1 . p(7)], t(2) |-> [1 . p(3),1 . p(5),1 . p(8)], t(3) |-> [2 . p(4),1 . p(6),
nilP], t(4) |-> [1 . p(6),2 . p(1),nilP], t(5) |-> [1 . p(0),1 . p(7),nilP])
1 . p(0) + 10 . p(2) + 10 . p(4)
\end{verbatim}
As you can see, the system enters a degraded configuration in which only line 1 is working and the raw pieces left on the faulty line (a half of the total) have been moved on line 1.
\begin{table}[t]
    \centering
\caption{Performance of \texttt{search} command as $M$ varies}
\begin{tabular}{ |p{1cm}||p{2cm}|p{2.5cm}|p{2cm}|  }
 \hline
 $M$ & \verb|#| states & \verb|#| rewrites & time (ms) \\
 \hline
 3  & 350      & 35730     &  44\\
 5  &  1232    & 130539    &  153\\
 10 & 8932     & 995724    &  1192\\
 20 & 84007    & 9737494   &  36090\\
 50 & 2186132  & 261564504 &  862536\\
 \hline
\end{tabular}
\label{tab:perf}
\end{table}

\subsection{Exploiting PT nets structural analysis}
In a somehow hybrid modelling framework like that proposed, we can benefit from the tools/techniques available for both formalisms.
For example, structural analysis of Petri nets may be an interesting, really efficient alternative/complement to state-space inspection techniques, when the system state-space is huge and performances degrade (as shown in Table \ref{tab:perf}).
Furthermore, structural analysis doesn't depend on the initial marking of a PT system, so we may use it to get some kind of parametric outcomes. Let us show an application of semiflows on this direction to our running example.   

Let $\mathbf{Q}$ be the $|P|\cdot |T|$ matrix such that is $\mathbf{Q}_{[p,t]} = 0(t)(p)-I(t)(p) \in \Rat$.
Any $P$-vector $\mathbf{p}$ which is a non-null, positive integer solution of the product $\mathbf{p} \cdot \mathbf{Q}=\mathbf{0}$, called $P$-semiflow, expresses a conservative law for the marking of places corresponding to non-zero entries of $\mathbf{p}$.
Any $T$-vector $\mathbf{t}$ which is a non-null, positive integer solution of the product $ \mathbf{Q}\cdot \mathbf{t}=\mathbf{0}$, called $T$-semiflow, expresses a cyclicality effect for firing sequences matching the semiflow.

By inspecting the Maude module \texttt{RWPT-FMS} one can (also formally) check that the system's structure may only be one of those described in Fig.~\ref{fig:FMS}, more a symmetrical faulty one.

Table~\ref{tab:semiflow} shows the semiflows\footnote{Semiflows have been computed with the \texttt{GreatSPN} tool \cite{GreatSPN}.} of the PT nets in Fig.~\ref{fig:FMS} (there are symmetric semiflows for the other faulty configuration). We observe that the PT nets are covered by $P$-semiflows, therefore, the whole system is structurally bounded. The $T$-semiflows represent a base production cycle in the nominal and faulty configurations of the MS. Consider the two $P$-semiflows of the faulty configuration, and the initial marking immediately after a system switch. We derive these invariant marking-expressions: $\forall m: m(p_0)+m(p_8) = 1$, $m(p_1)+2 \cdot m(p_6) + m(p_3) + m(p_5)= 2 \cdot M$. With simple arguments we can show that if $m(p_8) = 1$ then we \emph{eventually} reach either of the two dead states: $m' = 1 * p_8 + 2 \cdot M * p_3$, $m'' = 1 * p_8 + 1 * p_5 + (2 \cdot M - 1) * p_3$, corresponding to matches 1,4 found with the \texttt{search} command (for $M =50$). We have used this parametric outcome in rule \texttt{r2} of module \texttt{RWPT-FMS}. 

Even if, in general, a static prediction of all the possible structural changes of a rewritable PT system specified with Maude may be more complex, or even impossible, the opportunity to exploit the structural analysis capabilities of Petri nets together with the formal analysis tools of Maude looks promising. 

\begin{table}[t]
    \centering
\caption{P- and T-semiflows of the PT nets specifying the MS (Fig.~\ref{fig:FMS})}
\begin{tabular}{|c|c||c|c|}
  \hline
  \multicolumn{2}{|c||}{nominal behavior}&
  \multicolumn{2}{|c|}{faulty  behavior}\\
  \hline
  \hline
  \begin{tabular}{@{}l@{}} $\mathrm{pin_{1}}$\\ $\mathrm{pin_{2}}$\\ $\mathrm{pin_{3}}$\\
  $\mathrm{tin_{1}}$\\
  \end{tabular}&
  \begin{tabular}{@{}l@{}}
    \p{1} + 2*\p{6} + 2*\p{2} + 2*\p{4}\\
    \p{1} + 2*\p{6} + 2*\p{3} + 2*\p{5}\\
    \p{0} + \p{7} + \p{8}\\
    \tr{0} + \tr{1} + \tr{2} + \tr{3} + \tr{4}
  \end{tabular}&
  \begin{tabular}{@{}l@{}}
    $\mathrm{pin'_{1}}$\\
    $\mathrm{pin'_{2}}$\\
    \\
    $\mathrm{tin'_{1}}$\\
  \end{tabular}&
  \begin{tabular}{@{}l@{}}
    \p{1} + 2*\p{6} + \p{3} + \p{5}\\
    \p{0} + \p{8}\\
    \\
    2*\tr{0} + 2*\tr{2} + \tr{3} + \tr{4}
  \end{tabular}\\
  \hline

\end{tabular}    
    \label{tab:semiflow}
\end{table}

\section{Conclusion, open issues and ongoing work}
We have presented a Maude formalization of  ``rewritable'' PT nets, a framework for the specification/analysis of distributed system with dynamic reconfiguration capabilities.
With respect to similar approaches, the proposed encoding provides much more data abstraction to ease the modeller task, is more compact and efficient, and fosters the definition of rewrite rules with a high degree of flexibility. We have used as a (simple but tricky) running example throughout the paper a reconfigurable, fault-tolerant Manufacturing System. We have reported some experiments of formal verification of properties and discussed about the possible advantages of such a hybrid modelling approach.

\paragraph{Ongoing work \& open issues}
We first plan to enrich the modular and intuitive Maude specification with structural extensions (e.g., test/flush arcs, transition priorities) that further enhance the model's expressivity. The idea is to use ``decorated'' terms representing PT nodes (e,g, \verb|t(1,"line",0)|, where
there are a label and a value indicating the priority)
and to update the \texttt{firing} rule in module \texttt{PT-EMU} accordingly.

A more complex, really interesting extension is the Maude encoding of rewritable High-Level PN (e.g.,algebraic PN) possibly using the meta-level and term-unification modules.

The Labelled Transition System of a Maude specification of rewritable PT nets should be defined up to isomorphism of PT systems (in that case, the state-space of the running example should be reduced by half). A possible solution is to define a syntactical normal form for PT systems using equations. This is generally complex, but some heuristics could help dramatically reduce the inefficiency in most practical cases, e.g., by defining classes of "similar" PT nodes (using labels) so that an isomorphism preserves these symmetry classes. This topic, however, deserves special attention.

Rule-based transformations are simple to specify and elegant, but sometimes they are shown to be flawed and ill-equipped for representing realistic situations. Passing to more complex and suitable forms should be easy using the \emph{reflection} capability of Maude and/or the Maude \emph{strategy language} \cite{MARTIOLIET2009227} (a specification layer above those of equations and rules).

These advanced features of Maude might be also used to control the rewriting process and break the construction of the transition system of a rewritable PT net when an undefined/error state is reached. The strategy language provides a cleaner way, respecting the separation of concerns.

\bibliographystyle{eptcs}
\bibliography{biblio}

\newpage
\appendix
\section{Appendix - Maude specification of rewritable PT nets}
{\small
\begin{lstlisting}[language=maude]
fmod TRAN is
 protecting NAT .
 sort Tran .
 op t : Nat -> Tran [ctor] .
 op subscript : Tran -> Nat .
 vars N : Nat .
 eq subscript (t(N)) = N .
endfm

fmod IMATRIX is
 pr BAG{Place} * (sort Bag{Place} to BagP, sort NeBag{Place} to NeBagP, op nil to nilP) .
 pr EXT-BOOL .
 sort Imatrix .
 op [_,_,_] : BagP BagP BagP -> Imatrix [ctor] .
 op remove : Imatrix Place -> Imatrix .
 op in : Imatrix Place -> Bool .
 vars X Y Z : BagP .
 var P : Place .
 eq remove([X,Y,Z], P) = [set(X,P,0),set(Y,P,0),set(Z,P,0)] .
 eq in([X,Y,Z], P) = X[P] =/= 0 or-else Y[P] =/= 0 or-else Z[P] =/= 0 .
endfm

fmod PT-SYS is
 pr MAP+{Tran, Imatrix} * (sort Map{Tran, Imatrix} to Net, op empty to emptyN) .
 sort System .
 op __ : Net BagP -> [System] [ctor] .
 ops In Out Inh : Net Tran  -> [BagP] .
 op enabled : System Tran -> [Bool] .
 op dead : System -> Bool .
 op in : Net Place -> Bool . *** test the existence of a place
 var N  : Net .
 var T : Tran .
 var P : Place .
 var I O H S : BagP .
 var Q : Imatrix .
 var K K' : NzNat .
 eq In((T |-> [I,O,H], N), T) = I .
 eq Out((T |-> [I,O,H], N), T) = O .
 eq Inh((T |-> [I,O,H], N), T) = H .
 eq in((T |-> Q, N), P) = in(Q, P) or-else in(N, P). 
 eq in(emptyN, P) = false .
 eq T |-> [I,I,H] = emptyN [metadata "null t"] .  
ceq T |-> [K . P + I, O, K' . P + H] = emptyN if K >= K' [metadata "dead t"] .
*** ceq N K . P + S = N S if in(N, P) = false [metadata "isolated p"] .
 eq enabled( (T |-> [I,O,H], N) S, T) = I <= S and-then H >' S .
ceq dead((T |-> Q, N) S) = false if enabled( (T |-> Q, N) S, T) .
 eq dead(N S) = true [owise] .
cmb N S : System if N =/= emptyN . *** a PT system cannot be "empty"
endfm


fth PTSYSTH  is 
 protecting PT-SYS .
 op m0  : -> BagP .
 op net : -> Net .
endfth 


mod PT-EMU{X :: PTSYSTH} is
 var T : Tran .
 var I O H S : BagP .
 var N N' : Net .
 crl [firing] : N S => N S + O - I if T |-> [I,O,H], N' := N /\ I <= S /\ H >' S .
endm


mod RWPT-FMS is
 *** protecting PT-SYS .
 protecting PT-RWLIB .
 op net : -> Net .
 op m0  : -> BagP .  
 op M   : -> Nat . *** model's parameter
 vars N N' : Net .
 vars TL TF : Tran .
 vars P2 P3 P4 P5 PF : Place .
 var S : BagP .
 var K : NzNat .
 eq M = 50 . 
 eq net = t(0) |-> [2 . p(1), 1 . p(2) + 1 . p(3), nilP], t(1) |-> [1 . p(2) , 1 . p(4), 1 . p(7)],
    t(2) |-> [1 . p(3) , 1 . p(5), 1 . p(8)], t(3) |-> [1 . p(4) + 1 . p(5) , 1 . p(6), nilP], 
    t(4) |-> [1 . p(6) , 2 . p(1), nilP], t(5) |-> [1 . p(0) , 1 . p(7), nilP], t(6) |-> [1 . p(0) , 1 . p(8), nilP].
 eq m0  = 2 * M . p(1) +  1 . p(0) .
 
 crl [r1] : (N, t(0) |-> [2 . p(1), 1 . P2 + 1 . P3, nilP] , t(3) |-> [1 . P4 + 1 . P5, 1 . p(6), nilP],
   TF |-> [1 . p(0), 1 . PF, nilP], TL |-> [1 . P3, 1 . P5, 1 . PF]) S + 1 . PF => 
   (N, t(0) |-> [1 . p(1), 1 . P2, nilP],  t(3) |-> [2 . P4, 1 . p(6), nilP]) set(S, P3, 0) + S[P3] . P2 + 1 . p(0)
     if S[P5] = 0 .
 crl [r2] : N S   => net S + 1 . p(0) + M . P3 - M . P2 - 1 . p(7) - 1 . p(8)  if 1 . P2 := Out(N, t(0)) 
   /\  1 . P2 + 1 . P3 := Out(net, t(0))  /\ dead(N S) .
 *** crl [r3] : N S  => setw(N, t(0), p(9), i, 1) S if S[p(0)] = 0 /\ S[p(9)] = 0  /\ in(N, p(9)) = false .
endm

view Fms from PTSYSTH to RWPT-FMS is 
 op m0 to m0 .
 op net to net . 
endv

mod FMS-EMU is
 including PT-EMU{Fms} .
endm

fmod PT-RWLIB is
 protecting PT-SYS .
 sort Atype . *** arc type
 ops i o h : -> Atype [ctor] .
 op w : Net Tran Place Atype -> [Nat] . ***get an arc's weight
 op setw : Net Tran Place Atype Nat  -> Net . ***set an arc
 op setwS : Net Tran Place Atype Nat  -> Net . ***set an arc in a safe way
 ops remove : Place Net -> Net .
 var N N' : Net .
 var T : Tran .
 var P : Place .
 var I O H M : BagP .
 var Q : Imatrix .
 var K K' : NzNat .
 var Y : Nat .
 var A : Atype .
 eq w(N, T, P, i) =  In(N,T)[P] .
 eq w(N, T, P, o) = Out(N,T)[P] .
 eq w(N, T, P, h) = Inh(N,T)[P] .
 ceq setw(N, T, P, i, Y) =  N, T |-> [Y . P, nilP, nilP] if N[T] = undefined .
 eq setw((N, T |-> [I,O,H]), T, P, i, Y) =  N, T |-> [set(I, P, Y), O, H]  .
ceq setw(N, T, P, o, Y) =  N, T |-> [nilP, Y . P, nilP] if N[T] = undefined .
 eq setw((N, T |-> [I,O,H]), T, P, o, Y) =  N, T |-> [I, set(O, P, Y), H]  .
ceq setw(N, T, P, h, Y) =  N if N[T] = undefined .
 eq setw((N, T |-> [I,O,H]), T, P, h, Y) =  N, T |-> [I, O, set(H, P, Y)]  .
 eq remove(P, (T |-> Q, N) ) = T |-> remove(Q, P), remove(P, N) .
 eq remove(P, emptyN) = emptyN .
ceq setwS(N, T, P, A, Y) = N' if N' := setw(N, T, P, A, Y) /\ N' =/= emptyN .  
 eq setwS(N, T, P, A, Y) = N [owise] .
endfm
\end{lstlisting}
}

\section{Appendix - Maude generic modules}
{\small
\begin{lstlisting}[language=maude]
fmod BAG{X :: TRIV} is
 protecting INT .
 sorts Bag{X} NeBag{X} .
 subsorts NeBag{X} < Bag{X} .
 op nil : -> Bag{X} [ctor] .
 op _._  : Nat X$Elt ->  Bag{X} [prec 35 ctor] .
 op _._  : NzNat X$Elt ->  NeBag{X} [ctor ditto] .
 op _`[_`] : Bag{X} X$Elt -> Nat [prec 23] .
 op _+_  : Bag{X} Bag{X} -> Bag{X} [prec 39 ctor assoc comm id: nil] .
 op _+_  : NeBag{X} Bag{X} -> NeBag{X} [ctor ditto] .
 op _-_  : Bag{X} Bag{X} -> Bag{X} [prec 41 gather (E e)] .
 op _<=_ : Bag{X} Bag{X} -> Bool [prec 43] .
 op _>'_ : Bag{X} Bag{X} -> Bool [prec 43] .
 op set  : Bag{X} X$Elt Nat -> Bag{X} . 
 vars X Y : X$Elt .
 vars N M : NzNat .
 var K : Nat .
 var B B' : Bag{X} .
 eq (N . X + B)[X] = N .
 eq B[X] = 0 [owise] .
 
 eq 0 . X = nil .
 eq N . X + M . X = (N + M) . X .
 
 eq B - nil = B .
 eq N . X + B - M . X + B' = if N > M then N - M . X + (B - B') else B - B' fi  .
 eq B - M . X + B' = B - B' [owise] .
 
 ceq N . X + B <= B' = false if N > B'[X] .
 eq B <= B' = true [owise] . 
 
 ceq N . X + B >' B' = false if N <= B'[X] .
 eq B >' B' = true [owise] .
 
 eq set(B + N . X, X, K) = B + K . X . 
 eq set(B, X, K)  =  B + K . X  [owise] .
endfm

fmod MAP+{X :: TRIV, Y :: TRIV} is 
  sorts Entry{X,Y} Map{X,Y} .
  subsorts Entry{X,Y} < Map{X,Y} .
  op _|->_ : X$Elt Y$Elt -> Entry{X,Y} [ctor] .
  op empty : -> Map{X,Y} [ctor] .
  op _`,_ : [Map{X,Y}] [Map{X,Y}] -> [Map{X,Y}] [assoc comm id: empty ctor prec 121 ] .
  op undefined : -> [Y$Elt] [ctor] .
  op insert : X$Elt Y$Elt [Map{X,Y}] -> [Map{X,Y}] .
  op insert : X$Elt Y$Elt Map{X,Y} -> Map{X,Y} .
  op remove : X$Elt [Map{X,Y}] -> [Map{X,Y}] .
  op remove : X$Elt Map{X,Y} -> Map{X,Y} .
  op _`[_`] : [Map{X,Y}] X$Elt -> [Y$Elt] [prec 23] .
  op $hasMapping : [Map{X,Y}] X$Elt -> Bool .
  op hasduplicate : [Map{X,Y}] -> Bool .
  
  var D : X$Elt .
  vars R R' : Y$Elt .
  var M : [Map{X,Y}] .
  
  eq $hasMapping ((M, D |-> R), D) = true .
  eq $hasMapping (M, D) = false [owise] .
  eq insert (D, R, (M, D |-> R')) = if $hasMapping (M, D) then insert (D, R, M) else (M, D |-> R) fi .
  eq insert (D, R, M) = (M, D |-> R) [owise] .
  eq remove (D, (M, D |-> R)) = remove(D, M) .
  eq remove (D, M) = M [owise] .
  eq (M, D |-> R) [D] = if $hasMapping (M, D) then undefined else R fi .
  eq M [D] = undefined [owise] .
  eq hasduplicate((D |-> R, D |-> R', M)) = true .
  eq hasduplicate(M) = false [owise] .
 cmb M : Map{X,Y} if hasduplicate(M) = false .
endfm
\end{lstlisting}}

\end{document}